\documentclass[prb,twocolumn,showpacs]{revtex4-1}
\usepackage{amssymb,graphicx,color,amsmath,xfrac,bm}
\usepackage[all]{xy}

\renewcommand{\bold}[1]{{{\bm{#1}}}}
\renewcommand{\log}{{\mathrm{ln}}}
\renewcommand{\tanh}{{\mathrm{th}}}
\newcommand{\atanh}{{\mathrm{atanh}}}

\renewcommand{\Im}{{\mathrm{Im}}}
\renewcommand{\Re}{{\mathrm{Re}}}
\newcommand{\sign}{{\mathrm{sign}}}

\newcommand{\av}[1]{\ensuremath{\left\langle{#1}\right\rangle}}

\newcommand{\HT}[1]{{\mathrm{H.T.}\left\{#1\right\} }}
\newcommand{\mode}{{\mathcal{Q}}}
\newcommand{\qcf}{{{\mathcal{X}}}}
\newcommand{\raman}{{\chi}}
\newcommand{\Cutoff}{{\Lambda}}
\newcommand{\eps}{{\varepsilon}}
\newcommand{\eFermi}{{\epsilon_F}}
\newcommand{\smear}[1]{{\int{ds}\left(\frac{d\tanh\frac{#1 s}{2T}}{2ds}\right)}}
\newcommand{\SM}[2]{{ \wr\!\!\!\,\wr{ #2 }\wr\!\!\!\,\wr}_{#1}}   
\newcommand{\diagr}[1]{ \vcenter{\fancy{\xymatrix @-0.6pc @R=1.4pc  @M=0pt{ #1} }}}
\newcommand{\diagram}[2]{ \vcenter{\fancy{\xymatrix #1 { #2} }}}
\newcommand{\fancy}[1]{{\color{blue}#1}}
\newcommand{\intA}[1]{{\int\frac{\hat{\nu}(#1){d}{#1}}{2\pi}} }

\SelectTips{eu}{12}
\newcommand{\x}{\object@{>}}
\newcommand{\y}{\object@{<}}

\newcommand{\verfull}[5]{\fancy{{\vcenter{\xymatrix @-0.6pc @R=1.4pc  @M=0pt { 
\ar@{-}[r]|{\x}^{#1}&\ar@{-}[d]\ar@{}[dr]|{#3}\ar@{-}[r]&\ar@{-}[d]  \ar@{-}[r]|{\x}^{#4}&\\  
 &\ar@{-}[l]|{\x}^{#2} \ar@{-}[r] &  &  \ar@{-}[l]|{\x}^{#5} \\ }}}}}
\newcommand{\verpair}[2]{ \fancy{\vcenter{\xymatrix @-0.6pc @R=1.4pc  @M=0pt{  
\ar@{-}[r]|{\x}^{#1} &\\  &  \ar@{-}[l]|{\x}^{#2} \\ }}}}
\newcommand{\verbarestep}[3]{ \fancy{\vcenter{\xymatrix 
@-0.4pc @R=1.4pc  @M=0pt{
  &\ar@{~}[d]^{#1} \ar@{-}[r]|{\x}^<(.8){#2} &\\
   &  &  \ar@{-}[l]|{\x}^{#3} \\}}}}
\newcommand{\verbare}[5]{ \fancy{\vcenter{\xymatrix @-0.6pc @R=1.4pc  @M=0pt{  
\ar@{-}[r]|{\x}^{#1} &\ar@{~}[d]^{#3} \ar@{-}[r]|{\x}^<(.8){#4} &\\  &
\ar@{-}[l]|{\x}^{#2} & \ar@{-}[l]|{\x}^{#5} \\ }}}}
\newcommand{\verbaretot}[5]{ \fancy{ \vcenter{\xymatrix @-0.6pc @R=1.4pc  @M=0pt{  
\ar@{-}[r]|{\x}^{#1} &\ar@{-}@<-0.2pc>[d] \ar@{-}@<0.2pc>[d]^{#3} \ar@{-}[r]|{\x}^<(.8){#4}& \\ 
&\ar@{-}[l]|{\x}^{#2} & \ar@{-}[l]|{\x}^{#5}  }}}}
\newcommand{\versteptot}[8]{ \fancy{ \vcenter{\xymatrix  @-0.6pc @R=1.4pc  @M=0pt{
\ar@{-}[r]|{\x}^{#1}&\ar@{-}[d]\ar@{}[dr]|{#3}\ar@{-}[r] & \ar@{-}[d]  
\ar@{-}[r]|{\x}^{#4}& \ar@{-}@<-0.2pc>[d] \ar@{-}@<0.2pc>[d]^{#6} 
\ar@{-}[r]|{\x}^<(.8){#7} & 
\\ &\ar@{-}[l]|{\x}^{#2} \ar@{-}[r] &  &  \ar@{-}[l]|{\x}^{#5}  &  \ar@{-}[l]|{\x}^{#8}  }}}}
\newcommand{\verstep}[8]{  \fancy{\vcenter{\xymatrix  @-0.6pc @R=1.4pc  @M=0pt{
\ar@{-}[r]|{\x}^{#1}&\ar@{-}[d]\ar@{}[dr]|{#3}\ar@{-}[r] & \ar@{-}[d]  
\ar@{-}[r]|{\x}^{#4}&\ar@{~}[d]^{#6}\ar@{-}[r]|{\x}^<(.8){#7} & \\ 
 &\ar@{-}[l]|{\x}^{#2} \ar@{-}[r] &  &  \ar@{-}[l]|{\x}^{#5}  &  \ar@{-}[l]|{\x}^{#8} } }}}
\newcommand{\tadpole}[4]{{{ \fancy{\xymatrix @=0.7pc @M=0pt @C=1.2pc{
 \ar@{-}[r]|{\x}_{#1}  & \ar@/^1.1pc/@{~}[rr]^{#2} \ar@{-}[rr]|{\x}_{#3} &&  \ar@{-}[r]|{\x}_{#4}&}}}}}
\newcommand{\lefttriang}{{\fancy{\vcenter{ \xymatrix  @-1.5pc @R=.8pc  @M=0pt{ 
& \ar@{-}[dd] \\ \ar@{-}[ur] \ar@{-}[dr] & \\ & }}}}}
\newcommand{\righttriang}{{\fancy{\vcenter{\xymatrix  @-1.5pc @R=.8pc  @M=0pt{  
\ar@{-}[dd] & \\ & \ar@{-}[ul] \ar@{-}[dl] \\ & }}}}}
\begin{document}
\title{On the metallic state in cuprates.}
\author{Arkady~Shekhter}
\affiliation{National High Magnetic Field Lab, Tallahassee, Florida 32310}
\pacs{74.72.-h,78.30.Er, 67.10.Jn}
\begin{abstract}
We calculate Raman response functions on the Fermi surface in metallic cuprates. 
\end{abstract}
\date{\today}\maketitle

\section{Introduction.}

Cuprates in the metallic doping range have a large Fermi surface which is populated by short lived quasiparticles with lifetime strongly dependent on temperature. It has been proposed\cite{Varma1989} that short lifetime and strong temperature dependence are a result of coupling to fluctuations of a local charge state, which is independent of but coexists with the Fermi surface quasiparticles; to account for main experimental features the dynamics of the fluctuations of the charge state is to be local in space and non-local in time, $\mode(rt;r't')\propto\delta(r-r')\times[t-t']^{-1}$; this form of correlation function (dynamic mode) has recently been confirmed by microscopic calculation, $\Im\mode(\omega)\propto\tanh\omega/2T$; it obeys the $ \omega/T$ scaling, typical of the dynamics of fluctuations near the quantum critical point.\cite{VarmaAji} On the microscopic level, the local charge state is a pattern of inter-atomic currents within copper-oxide unit cell\cite{VarmaLoop} (loop currents) which has a point group symmetry equivalent to a polar vector parallel to the copper-oxide plane, a toroidal moment\cite{SV}. Once non-zero, the only dynamics possible for a local toroidal moment is to rotate its orientation. The part of the Hamiltonian which describes coupling between the Fermi surface quasiparticles and the local charge state has a form  $\mathcal{H}_{coupling}\propto U(\bold{r})[\bold{\nabla}\times\bold{j}(\bold{r})]$. Here operator $U(\bold{r})$ acts on the charge state; it generates rotation of the local toroidal moment around the axis perpendicular to copper-oxide plane\cite{ASV}. The Fermi surface operator $[\bold{\nabla}\times\bold{j}(\bold{r})]$ (where  $\bold{j}(\bold{r})$ is quasiparticle current) has a symmetry of angular momentum perpendicular to the  copper-oxide plane. Using a language of group theory, the dynamic mode $\mode(\omega)$ couples to Fermi surface quasiparticles via an operator which transforms under $A_{2g}$ representation of the lattice point group, the same representation to which $z$-axis component of the angular momenta belongs. 

Raman scattering directly probes charge dynamics in the system; the Raman cross-section $d\sigma \propto [1+n_B(\omega)]\Im\raman_{\alpha}(\omega)$ is proportional to the Raman response function, $\raman_{\alpha}(\omega) \equiv \raman_{\alpha}(\omega,q=0)=\langle\!\langle\rho_{\alpha}\rho_{\alpha}\rangle\!\rangle_{\omega}$  where $\rho_{\alpha}$ is charge density operator in the  lattice symmetry channel  $\alpha$. In cuprates (which typically have tetragonal lattice) the cartesian product of the polarization vectors of incoming and outgoing photons breaks into irreducible representations  $\alpha=A_{1g},A_{2g},B_{1g},B_{2g},E_g$,  see, e.g., Ref.~\onlinecite{Devereaux}. These are symmetry channels for which experimental data for Raman scattering in cuprates is specified. 
Naturally, Raman signal in  $A_{2g}$ symmetry channel would probe directly dynamics of the local charge state and reveal the  $\omega/T$ scaling, $\chi(\omega)_{\alpha=A_{2g}}\propto \tanh\omega/2T$. It is known that Raman spectra with similar behavior is observed\cite{Raman} in symmetry channels, other than $A_{2g}$. In this paper we demonstrate that Raman response in symmetry channels such as $B_{1g},B_{2g}$ originates on the Fermi surface; we show that particle-hole response exhibits  $\omega/T$ scaling if quasiparticle coupling to the dynamic mode is allowed. The experimental situation will be addressed in a separate publication. 

The response functions are analyzed in terms of dynamics of excitation states; in metals near the Fermi surface these are particle-hole pairs. In conventional metals the dynamic process is particle-hole rescattering mediated by screened coulomb interaction\cite{Landau}. In cuprates interaction with the dynamic fluctuation has two major effects on the quasiparticle physics. First, charge fluctuations serve as a time-dependent environment which is capable of exchanging energy with the quasiparticle; as a result, quasiparticle lifetime becomes short and temperature dependent. Second, dynamic fluctuation can mediate particle-hole rescattering; unlike conventional metals where the particle-hole rescattering event is local in time, $\delta t\sim{\hbar}/\eFermi$, in cuprates rescattering event can be highly non-local in time. 

Similar situation arises in many other strongly correlated metals\cite{Herring,Stewart,VarmaMixed,VarmaReview}. In this paper we focus on cuprates in the ``strange metal'' phase\cite{Varma1989,Abrahams1996,Kotliar1991}, the region of phase diagram in the metallic doping range at temperatures above superconducting and pseudogap transitions, though we expect that our discussion is applicable more generally. 

\section{response functions on the Fermi surface in the metallic cuprates.}

The dynamic mode in cuprates in the metallic doping range is specified by its (retarded) correlation function\cite{Varma1989,VarmaAji,ASV}
\begin{multline}\label{eq:MFL-chi}
\diagram{@+0.1pc @R=1.5pc  @M=0.1pt}{\ar@{~}[r]|{\x}^{\omega,q} &}
= \qcf_0\mode(\omega,q)^R = \langle\!\langle U(rt)U^{\dagger}(r't')\rangle\!\rangle_{\omega,q} \\
\Im \mode(\omega,q)^R = \begin{cases}
-\tanh\frac{\omega}{2T}, & \omega<\Cutoff_0\\
0, & \omega>\Cutoff_0
\end{cases}
\end{multline}
where $U(rt)$ operates on the local charge state; it generates rotation of the local toroidal moment (loop current state), see Ref.~\onlinecite{ASV} for details. The cutoff $\Cutoff_0$ is determined from ARPES data\cite{ARPES} and is found to be  $\sim0.4eV$. The spectral weight $\qcf_0$ can be estimated from the results of polarized neutron experiments\cite{Neutron} Coupling of the dynamic fluctuation to the Fermi surface quasiparticles is specified in terms of the vertex 
\begin{align}\label{eq:couplingfunction}
&\diagr{  
& &\\
&\ar@{~}[l]|{\y}^{} \ar@{-}[ru]|{\x}^{\bold{k}} \ar@{-}[rd]|{\y}_{\bold{k}'}  & \\
& & }
= \gamma_0 U_{k-k'}\; \Big[ \phi(\bold{k},\bold{k}')  c^{\dagger}_{\bold{k}} c_{\bold{k}'} \Big] , 
\end{align}
where $\gamma_0$ is the coupling constant; operators $c_{\bm{k}}$ are fermion quasiparticle operators on the Fermi surface. Function $\phi(\bold{k},\bold{k}')$ is a normalized angular harmonics on the lattice which transforms under $A_{2g}$ irreducible representation of the $D_{4h}$ group. From the space-group symmetry perspective it behaves as $i \bold{k}\times\bold{k}'$; see Ref.~\onlinecite{ASV} for more details and derivation of Eq.~(\ref{eq:couplingfunction}). The elementary particle-hole rescattering event mediated by the exchange of the dynamic mode is specified with  
\begin{align}\label{eq:rescatteringvertex}
\diagram{@+0.1pc @R=1.5pc  @M=0.1pt}{  
& \ar@{~}[d]|{\y}^{}  
\ar@{-}[l]|{\y}_<(0.8){\eps,\bold{k}} \ar@{-}[r]|{\x}^<(0.8){\eps',\bold{k}'} & \\
&\ar@{-}[l]|{\x}^<(0.8){\bold{k}} \ar@{-}[r]|{\y}_<(0.8){\bold{k}'}&}
\!\!\!\!\!\!\!\!\!\!=g(\eps-\eps'; k,k') = \gamma_0^2\qcf_0 \nu \mode(\eps-\eps')|\phi(\bold{k},\bold{k}')|^2
\,,
\end{align}
where $\nu$ is density of states on the Fermi surface (per one spin species). We define lattice angular harmonics $F_{\alpha}(\hat{k})$ via the decomposition 
\begin{align}\label{eq:aaa}
g(\eps-\eps'; k,k') = \sum g_{\alpha}\mode(\eps-\eps')F_{\alpha}(\hat{k}) F_{\alpha}(\hat{k}') 
\end{align}
The coupling constant $\gamma_0$, the spectral weight $\qcf_0$ and the Fermi surface density of states $\nu$  are absorbed in the definition of dimensionless coupling constants $g_{\alpha}$  and do not appear explicitly in the calculation and results. In cuprates the density of states on the Fermi surface is non-uniform. The momentum integration is performed with 
\begin{align}\label{eq:aaa}
\int\frac{d^2p'}{(2\pi)^2} \cdots = {\nu_0}\int d\xi\int\frac{\nu(\theta)/\nu_0}{2\pi} \cdots  
\end{align}
where function $\nu(\theta)$ has the symmetry of the lattice, i.e., it belongs to a scalar representation $A_{1g}$. ; in the following we use notation $\hat{\nu}(\theta)\equiv \nu(\theta)/\nu_0$ where $\nu_0 = \int(d\theta/2\pi) \nu(\theta)$ is average density of states. The angular harmonics  $F_{\alpha}(\hat{k})$ are normalized with~:
\begin{align}\label{eq:aaa}
\int \frac{d\theta}{2\pi} \hat{\nu}(\theta) F_{\alpha}(\theta) F_{\beta}(\theta) = \delta_{\alpha\beta}
\end{align}
Analysis of ARPES data\cite{ARPES} indicates that the coupling constant in the density, $\alpha=A_{1g}$, channel (which controls the inelastic part of quasiparticle selfenergy) is of order of unity; we will use notation $g_0=g_{A_{1g}}$. On symmetry grounds, within the point group $D_{4h}$ of the tetragonal lattice the angular decomposition of  $|\phi(\bold{k},\bold{k}')|^2$ in Eq.~(\ref{eq:rescatteringvertex})  contains only three irreducible representations\cite{Birss}; these are symmetry channels $B_{1g}$ , $B_{2g}$ and $A_{1g}$. For spherical Fermi surface $g_{B_{1g}}/g_{A_{1g}}=g_{B_{2g}}/g_{A_{1g}}=-1/2$; for Fermi surface in cuprates these dimensionless parameters can be determined from the experiment. We note that decoupling of the energy and momentum dependences in $g(\eps-\eps'; k,k')$ is a consequence of the fact that in cuprates the correlation function of the dynamic fluctuation  $\mode(\omega,q)$ is local in space.\cite{Varma1989,VarmaAji} 

In conventional metals the low-lying excited states are represented by a collection of particle-hole pairs on top of the Fermi surface.  Response functions are analyzed in terms of particle-hole rescattering which in mathematical description is represented by particle-hole ladder diagram. Free propagation of particle-hole pair, represented by the particle-hole section, is interrupted by a rescattering events mediated by Coulomb interaction and represented by Landau parameters. An observation which is a basis for the whole picture is that particle-hole section is singular near the Fermi surface; the form of singularity is determined by continuity equation which follows from particle number conservation property.\cite{Landau} By iterating particle-hole sections the particle-hole ladder collects (and exhaust) all singular contributions. This way the mathematical formalism does not make use of any small parameter; it only relies on the long quasiparticle lifetime which is necessary for developing a singular particle-hole section near the Fermi surface. In conventional metal the only channel for relaxation of quasiparticle energy is by emitting particle-hole pairs which are rare close to the Fermi surface because of the available phase space is small. In the metallic state of cuprates there is a new channel for quasiparticle energy relaxation. Emission and absorption of dynamic fluctuation is not confined by any phase space restrictions; in result there is a large inelastic quasiparticle relaxation rate near the Fermi surface and, consequently, short inelastic lifetime.\cite{Varma1989,Abrahams1996,Kotliar1991}

We base our discussion of the response functions in the metallic state in cuprates on the assumption that even though the particle-hole section has weaker singularity (branch cut, not a pole), it is still singular enough to justify the use of particle-hole ladder near the Fermi surface. Physically, this corresponds to analysis in terms of particle-hole rescattering. The technical side of the problem is more complex here compared to conventional metal. Indeed, particle-hole rescattering mediated by exchange of dynamic fluctuation is non-local in time, $\mode(tt')\propto1/t-t'$. In the mathematical analysis we make sure that particle number conservation is obeyed. The calculation is controlled by maintaining the balance between an inelastic quasiparticle relaxation rate (``outs'') and particle-hole rescattering (``ins'')\cite{Landau,Leggett,Baym,Eliashberg}. In that sense the present analysis follows the spirit of the conventional Fermi liquid theory. 

Particle-hole rescattering physics near the Fermi surface determines only the dynamic part of the response functions.\cite{Leggett,MFL} This is enough for the purpose of calculating the Raman response which requires calculation of $\Im\raman_{\alpha}(\omega)$; the static part of $\raman_{\alpha}(\omega)$ is real. Complete particle-hole rescattering process consists of an arbitrary sequence of two kinds of rescattering events~: those which are local in time (represented by Landau parameter  $B^{\alpha}$)  and those which are non-local in time. In the particle-hole ladder we identify segments $Z_{\alpha}(\omega)$ (singular segment) which describe sequence of particle-hole rescattering mediated by dynamic mode only. The correlation function $\raman_{\alpha}(\omega)$ has a structure of a geometric series in which there is an arbitrary number of singular segments interrupted by the Landau amplitude $B_{\alpha}$
\begin{align}\label{eq:RamanCF}
&\raman_{\alpha}^{\text{dyn}}(\omega) = 
\phi_{\alpha} \frac{ Z^{\alpha}_{\omega} }{ 1 - B_{\alpha}Z^{\alpha}_{\omega} }\phi_{\alpha}\,.
\end{align}
Here $\phi_{\alpha}$ is static Raman vertex (which also includes static Fermi liquid renormalizations). Landau amplitudes are defined via the angular decomposition of the non-singular scattering amplitudes $B(\hat{k},\hat{k}') = \sum B_{\alpha} F_{\alpha}(\theta)^B F_{\alpha}(\theta')^B$; here $F_{\alpha}(\theta)^B$ are normalized angular harmonics. Index $B$ indicates that these are different from angular harmonics $F_{\alpha}(\theta)$; in the same representation their overlap is not equal one, $\int (d\theta/2\pi) \hat{\nu}(\theta) F_{\alpha}(\theta) F_{\beta}(\theta)^B = f_{\alpha}\delta_{\alpha\beta}$, though they are orthogonal when in different representations. 

Analysis of particle-hole rescattering within the singular segment leads to the following expression 
\begin{multline}\label{eq:EquationZangles}
\bm{Z_{\omega}}^{\alpha} =-\int{d}\eps\frac12[\tanh\frac{\eps+\omega}2-\tanh\frac{\eps}2]
\\ 
\times\intA{\theta} 
F_{\alpha}(\theta)^B\,\bm{Y_{\omega}}^{\alpha}(\eps)_{\theta}\,F_{\alpha}(\theta)^B
\end{multline}
where the function $\bm{Y}(\epsilon)$ is determined from the equation  
\begin{multline}\label{eq:EquationYangles}
\omega \bm{Y_{\omega}}(\eps)^{\alpha}_{\theta} = 1\\
+\int d\eps' \mode^K_{\omega}(\eps'-\eps)\intA{\theta'}\\
\times\Big[g_{\alpha}F_{\alpha}(\theta)F_{\alpha}(\theta') \bm{Y_{\omega}}^{\alpha}(\eps')_{\theta'} - g_{0}F_{0}(\theta)F_{0}(\theta') \bm{Y_{\omega}}^{\alpha}(\eps)_{\theta}\Big]
\end{multline}
where kernel $\mode^K_{\omega}(\eps'-\eps)$ is defined in Eq.~(\ref{eq:Tkernel}); see next section for details. The singular segment is interrupted by non-singular rescattering event at both ends, hence the angular harmonics $F^{\alpha}(\theta)^B$ in Eq.~(\ref{eq:EquationZangles}). In the density channel, $\alpha=A_{1g}$, the solution of Eq.~(\ref{eq:EquationZangles}) is $\bm{Y}_{\omega}(\epsilon)=1/\omega$  and consequently $\bm{Z}_{\omega}^{A_{1g}}=-1$. This ensures particle conservation; see detailed discussion in Section~\ref{sec:conservation}. Eq.~(\ref{eq:EquationY}) has a form reminiscent of the kinetic equation, the integral on the right hand side plays a role of the collision integral. It is shown in the next section that the term $-g_{0}Y_{\omega}^{\alpha} (\eps)$ originates in the inelastic quasiparticle scattering rate (``outs'') and the term $g_{\alpha}Y_{\omega}^{\alpha} (\eps)$  originates in the process of particle-hole rescattering (``ins''). Eq.~(\ref{eq:EquationY}) replaces kinetic equation in the metallic state of cuprates;  we have to restore the  $q$-dependent term in the left-hand side, account for elastic disorder and local in time rescattering events, and replace the trigonometric factors in the definition of $\mode^K_{\omega}$ by the fermion and boson distribution functions. In this form $Y_{\omega,q}(\eps)$ now depends on $q$ and is suitable for the analysis of transport properties; these however require separate investigation. We note that the structure of Eqs.~(\ref{eq:EquationY},\ref{eq:EquationZ}) is independent of the particular form of correlation function of the dynamic mode. They are valid for a description of the particle-hole rescattering near the Fermi surface in the presence of dynamic mode of arbitrary form if validity of particle-hole approximation has been established.

Analysis of Eq.~(\ref{eq:EquationYangles}) in Raman channels leads to the result 
\begin{align}\label{eq:RamanCFall}
&\raman_{\alpha}^{dyn}(\omega)_T = 
(\phi_{\alpha})^2 \frac{ -\omega }{ (1+B_{\alpha})\omega + 2(g_0-g_{\alpha})\Psi(\omega)_T} 
\end{align}
The function $\Psi(\omega)$ has the same analytic properties as quasiparticle selfenergy; it has a form 
\begin{multline}\label{eq:resultZ}
\Psi^{\alpha}(\omega)_{T} = \smear{p} \Psi^{\alpha}(\omega-s)_{T=0}\\
\begin{aligned}
\Im\Psi^{\alpha}(\omega)_{T=0} =& |\omega|, \\
\Re\Psi^{\alpha}(\omega)_{T=0} =&  \omega\log\frac{\Cutoff_{\alpha}}{|\omega|} 
\end{aligned}
\end{multline}
where $p\approx0.59$. The cutoff  $\Cutoff_{\alpha}$ cannot be determined within the present calculation because it is not determined by the physics in the vicinity of the Fermi surface; it is reasonable to expect that $\Cutoff_{\alpha}$ is of order of the high energy cutoff in the fluctuation mode, $\Cutoff_0$. Note the factor $g_{\alpha}-g_0$. It is a result of partial cancellation between the dynamic singularity in the selfenergy ($\propto g_0$) and the singularity introduced by non-local in time particle-hole rescattering, both are due to the coupling of the Fermi surface quasiparticles with the dynamic mode. The analyticity of $\raman_{\alpha}(\omega)$ in the upper half of the complex plane is guaranteed by the fact that $g_{\alpha}-g_0>0$. The imaginary part of response function in Raman channels $\alpha\neq A_{1g}$ calculated from Eq.~(\ref{eq:RamanCFall}) obeys  $\propto\omega/T$ scaling at $\omega\lesssim{T}$; at large energy,  $\omega\gtrsim{T}$, it levels to a constant. 

In cuprates the dynamic mode can only induce particle-hole rescattering in ``tensor'' symmetry channels, $A_{1g},B_{1g},B_{2g}$; it does not participate in the particle-hole rescattering in the ``vector'' channels, such as $E_{g}$ and $E_{u}$. We now briefly discuss consequences of this fact. The current vertex (charge, heat, etc.) is spatial vector and it belongs to $E_u$ irreducible representation. Therefore in the direct calculation of the diagram for conductivity tensor the dynamic mode enters only via selfenergy; no particle-hole rescattering mediated by the mode is possible. This justifies single-bubble calculation of conductivity. The case of long-range disorder is captured by elastic disorder ladder in the transport diagram; this is equivalent to the discussion in terms of conventional kinetic equation\cite{HallVarmaAbrahams}. An example of the transport property where particle-hole rescattering via dynamic fluctuation is important is Fermi surface contribution to the solid viscosity\cite{Kinetics} which enters in the acoustic sound attenuation. Indeed, the stress tensor in the tetragonal lattice breaks into $A_{1g},B_{1g},B_{2g},E_g$ irreducible pieces. The solid viscosity is determined by the dynamic stress-tensor response function. We note that, unlike Raman correlation functions for which  $qv_F\ll\omega$, in acoustic sound the relation between frequency and momenta is reverse, $qv_F\gg\omega$; this requires analysis of q-dependent kinetic equation; see discussed after Eq.~(\ref{eq:EquationYangles}).

\section{particle-hole rescattering mediated by dynamic fluctuation}

\subsection{quasiparticle selfenergy}

In the remainder of this paper we indicate sequence of mathematical steps necessary to arrive to Eqs.~(\ref{eq:EquationYangles},\ref{eq:EquationZangles},\ref{eq:RamanCFall}). We use temperature technique which allows direct access to analytic structure of physical objects at hand; for details on analytic continuation in temperature diagrams see, e.g., Ref.~\onlinecite{Pitaevskii}. To make discussion more readable we assume the Fermi surface is spherical; angular harmonics can be easily restored as is discussed above. 

Emission and absorption of the dynamic mode by the quasiparticles on the Fermi surface results in an inelastic relaxation rate which is analyzed in terms of the selfenergy~:
\begin{align}\label{eq:first}
\Sigma(i\eps_n) 
=&  \tadpole{i\eps; p}{i\omega; p-p'}{i\eps-i\omega; p'}{i\eps; p} \notag\\ 
=& -g_0 T\sum_{i\omega_n}\int\frac{d^2p'}{\nu_0(2\pi)^2} G(i\eps_n-i\omega_n) \mode(i\omega_n)
\end{align}
Making analytic continuation we obtain
\begin{align}
&\tadpole{i\eps}{i\omega}{i\eps-i\omega}{i\eps}
\xrightarrow{i\eps_n>0} 
-g_0\int\frac{d\omega}{4\pi i} \int \frac{d^2p'}{\nu_0(2\pi)^2} \notag\\  
&\left[ 
\coth\frac{\omega}{2T} \;\tadpole{R}{R-A}{R}{R} 
-\tanh\frac{\omega-\eps}{2T} \;\tadpole{R}{R}{R-A}{R} \right]
\end{align}
or 
\begin{multline}\label{eq:second}
\Im\Sigma^R(\eps,p) =
-g_0 \int\frac{d\omega}{2\pi}\int\frac{d^2p'}{\nu_0(2\pi)^2} \\
\times[\coth\frac{\omega}{2T}-\tanh\frac{\omega-\eps}{2T}]  
\Im G^R(\eps-\omega,p')\Im\mode(\omega) 
\end{multline}
Let us comment on the higher order terms in the selfenergy. 
\begin{align}\label{eq:higherterms}
\vcenter{\fancy{\xymatrix @=1pc @M=0pt @C=1pc{
\ar@{-}[r]|{\x}_{} &\ar@{-}[r]|{\x}_{} \ar@/^1.8pc/@{~}[rrrr]^{} 
&\ar@/^1.1pc/@{~}[rr]^{} \ar@{-}[rr]|{\x}_{} &&\ar@{-}[r]|{\x}_{}&\ar@{-}[r]|{\x}_{}  \ar@{-}[r]|{\x}_{}&}}}
\end{align}
Making an analytic continuation in this diagram we find that all pairs of Green's functions, except the middle line, are $G^RG^R$ or $G^AG^A$ pairs. The energy integration is not confined to the Fermi surface here; naive momentum integration for the pair of Green's functions gives zero since in the complex plane of $\xi$ both poles are on the same side and expression falls fast enough $\sim1/\xi^2$ at large $\xi$.  However, this contribution is not zero: as noted above the frequency integral is not confined to the Fermi surface in RR section; the combined $d\eps d\xi/[(\eps-\omega-\xi)(\eps-\xi)]$ integral is logarithmically divergent at the upper limit and as usual have to be cut off at the Fermi energy scale. However, we observe that this contribution does not lead to a singular behavior in $\eps$ (or an imaginary part). We only collect the singular (non-analytic) at the Fermi surface terms. 

Momentum integration is trivial in Eq.~(\ref{eq:second}) owing to the fact that the mode is local in space; we use an identity 
\begin{align}\label{eq:aaa}
\int \frac{d^2p'}{(2\pi)^2} \Im G^R(\eps, p) = \nu_0\int_{-\infty}^{\infty} d\xi \Im G^R(\eps,\xi) = -\nu_0\pi
\end{align}
and obtain
\begin{align}\label{eq:Im-Sigma}
\Im\Sigma^R(\eps) =& g_0 \int d\omega \frac12 [\coth\frac{\omega}{2T}-\tanh\frac{\omega-\eps}{2T}]  
\;\Im\mode^R(\omega) \,.
\end{align}
The real part is calculated from Kramers-Kronig relations
\begin{align}\label{eq:aaa}
	\Re f(\omega) = \HT{\Im f(\omega)} \equiv \frac1{\pi}\; P\!\!\!\!\!\!\int d\omega'
        \frac{ \Im f(\omega')}{\omega'-\omega}\,.
\end{align}
We use Eq.~(\ref{eq:MFL-chi}) in Eq.~(\ref{eq:Im-Sigma})  and obtain (at $\eps\ll \Cutoff_0)$ 
\begin{align}\label{eq:ImSigma-exact}
\Im\Sigma(\eps) =& -g_0 \frac{\eps}{\tanh[\eps/2T]}\notag\\
\Re\Sigma(\eps) =& -g_0 \HT{\frac{\eps}{\tanh[\eps/2T]}}
\end{align}
To analyze this expression we introduce a representation which express the finite-temperature self-energy in terms of its form at zero temperature. To illustrate the idea we first calculate $\mode(\omega)_T$. Introduce a ``temperature smearing'' operator\cite{SF}, 
\begin{align}\label{eq:Smearingdef}
g(\omega) = \SM{p}{f(\omega)} \equiv\smear{p}{ f(\omega-s)}  
\end{align}
Starting with  
\begin{align}\label{eq:aaa}
\Im\mode(\omega)_{T=0} = -\sign\omega
\end{align}
and using an identity $\tanh({\omega}/{2T}) = \SM{p=1}{\sign\omega}$, we write the imaginary part of the selfenergy, Eq.~(\ref{eq:MFL-chi}), in the form    
\begin{align}\label{eq:aaa}
\Im \mode^R(\omega)_T = \SM{p=1}{ \Im\mode^R(\omega)_{T=0}} 
\end{align}
An important property of the operator $\SM{p}{\cdots}$  is that it commutes with Hilbert transform if we restrict ourselves to functions which are analytic in the upper/lower half of the complex plane. Indeed, smearing operation represents function $g(\omega)$ as a linear combination of functions $f(\omega)$ shifted parallel to the real axis. If function $f(\omega)$ is analytic in the upper half of the complex plane, so is the linear combination in the right-hand side of Eq.~(\ref{eq:Smearingdef}). As a consequence, both sides in Eq.~(\ref{eq:Smearingdef}) satisfy Kramers-Kronig relation with branch cut along the real axis. We immediately obtain 
\begin{align}\label{eq:ReChiFiniteT}
\Re\mode(\omega)_{T} = \SM{p=1}{\Re\mode(\omega)_{T=0}} 
\end{align}
and therefore 
\begin{align}\label{eq:SmearingChi}
\mode(\omega)_T=\smear{}\mode(\omega-s)_{T=0}
\end{align}
 At $T=0$ one have 
\begin{align}\label{eq:aaa}
\Re\mode(\omega)_{T=0} = \HT{\Re\mode(\omega)_{T=0}} = -\frac2{\pi}\log\frac{\Cutoff_0}{|\omega|} 
\end{align} 
Using smearing representation in Eq.~(\ref{eq:SmearingChi}) we write
\begin{align}\label{eq:Re-chi-smear}
\Re\mode^R(\omega)_T =& \smear{} \left[-\frac2{\pi} \log\frac{\Cutoff_0}{|\omega-s|}\right]
\end{align}
Expression in Eq.~(\ref{eq:Re-chi-smear}) gives a very intuitive representation for $\mode(\omega)_T$ at finite temperature. It makes it transparent that when $\omega\gg T$ the difference between  $\mode(\omega)_T$ and $\mode(\omega)_{T=0}$ is small. It also isolates the logarithmic divergence in the Hilbert transform which exists already at zero temperature. At small $\omega\ll T$ the integral in the real part can be evaluated in closed form using  $\int  ds\left(\frac{d\tanh(ps)}{2ds}\right) \log|s| = -\gamma-\log(4p/\pi)$; real part tends to a finite limit as $\omega \rightarrow0$. We can write an interpolating formula
\begin{align}\label{eq:aaa}
\Re \mode^R(\omega) \approx -\frac2{\pi}\,\,\log\frac{\Cutoff_0}{\max\{\omega,C T\}} 
\end{align}
where $C=\pi e^{-\gamma}/2\approx0.88$ and $\gamma\approx0.57$ is Euler's gamma. 

We now obtain similar ``smearing'' representation for the quasiparticle selfenergy at finite temperature. At zero temperature the selfenergy, Eq.~(\ref{eq:Im-Sigma}), is easily evaluated 
\begin{align}\label{eq:ImSigmaZeroT}
&\Im \Sigma^R (\eps)_{T=0} = g_0\int_0^{\eps}d\omega\Im\mode^R(\omega)_{T=0}
= - g_0\min( |\eps|, \Cutoff_0 )
\end{align}
The real part is 
\begin{multline}\label{eq:aaa}
\Re\Sigma(\eps)_{T=0} = \HT{\Im \Sigma(\eps)_{T=0}}\\
=- g_0\frac2{\pi} \Big[\!
	\eps\ln\!\frac{\Cutoff_0}{\eps}\!+\!
	\frac12(\eps\!+\!\Cutoff_0)\ln\!\frac{\eps\!+\!\Cutoff_0}{\Cutoff_0}\!+\! 
	\frac12(\eps\!-\!\Cutoff_0)\ln\!\frac{\eps\!-\!\Cutoff_0}{\Cutoff_0}
	\!\Big]\\
\xrightarrow{\eps\ll\Cutoff_0} 
-g_0\frac2{\pi}\big[\eps\ln\frac{\Cutoff_0}{|\eps|} +\eps\big]
\end{multline}
Following the analogous discussion for the dynamic mode $\mode(\omega)$ we first find the smearing representation for the imaginary part. Using an identity 
\begin{align} \label{eq:aaa}
\int ds \frac{d\tanh{\frac{p s}2}}{2ds} {|\omega-s|} 
= \frac2p \log\Big[ 2\cosh\frac{p\omega}2\Big]
\end{align}
and comparing with the finite temperature expression for the self-energy, Eq.~(\ref{eq:ImSigma-exact}), we write   
\begin{align}\label{eq:Interpolation-ImSigma}
\Im \Sigma^R (\eps) \approx & -g_0  \frac{2T}p \log\left[2\cosh\frac{p \eps}{2T} \right]  
\end{align}
For any value of $p$ this expression has correct behavior at $\eps\gg T$. The choice of $p$ is determined by the form of selfenergy at small $\eps\lesssim T$. One choice is $p=\ln2\approx0.69$ which ensures correct value at $\omega=0$.  The other choice is $p=2/3\approx 0.66$ which ensures correct value of the curvature at $\eps=0$. Two values of $p$ are close to each other and indeed for any value of $p$ between these two the difference between two sides of Eq.~(\ref{eq:Interpolation-ImSigma}) is barely visible on the plot. Eq.~(\ref{eq:Interpolation-ImSigma}) provides a good interpolation in the whole range of $\eps$. We conclude
\begin{align}\label{eq:good-interpolation}
\Im \Sigma(\omega)_T = \SM{p}{ \Im\Sigma(\omega)_{T=0} }
\end{align}
and therefore
\begin{align}\label{eq:SmearingSigma}
\Sigma(\omega)_T = \smear{p}\Sigma(\omega-s)_{T=0}
\end{align}
with $p\approx\ln2$. Note that $p=1$ in Eq.~(\ref{eq:SmearingChi}).

Finally, we indicate how these results are modified when density of states is non-uniform on the Fermi surface. Eq.~(\ref{eq:second}) is replaced with 
\begin{multline}\label{eq:Sigma-angles}
\Im\Sigma(\eps)_{\theta}^R =
-g_0 F_{0}(\theta) \int\frac{d\omega}{2\pi}\int\frac{d^2p'}{(2\pi)^2} F_{0}(\theta')\\
\times[\coth\frac{\omega}{2T}-\tanh\frac{\omega-\eps}{2T}]  
\Im G^R(\eps-\omega,p')\Im\mode(\omega) \\
\shoveleft{=-g_0F_{0}(\theta)\av{F_0}\Im\Sigma(\eps)^R }
\end{multline}
where $\Im\Sigma(\eps)^R$ is given by Eq.~(\ref{eq:Im-Sigma}). Here $\av{F_0}=\int d\theta(2\pi)\hat{\nu}(\theta)F_0(\theta)$. For non-circular Fermi surface the constant $g_0\av{F_0}$ plays the role of the dimensionless coupling constant in the density channel. The selfenergy acquires slow dependence on the position on the Fermi surface\cite{ZhuVarma}.

\subsection{Calculation of singular segment $Z_{\omega}^{\alpha}$.}

The singular segment $Z^{\alpha}_{\omega}$ describes multiple rescattering of particle-hole pairs via an exchange of a dynamic mode 
\begin{align}\label{eq:Zalpha}
Z^{\alpha}_{\omega} = \lefttriang\verpair{}{} \righttriang + \lefttriang\verpair{}{}\!\!\!\!\!\!\!\!\!\!\!\!\verbarestep{}{}{} \righttriang  + \cdots 
\end{align}
With our definition of the singular segment, Eq.~(\ref{eq:RamanCF}), the triangles in this expression correspond to a factor of one. In Ref.~\onlinecite{MFL} $Z_{0}(\omega)$ has been expressed in terms of a fully renormalized particle-hole propagation amplitude  $\Gamma^{I/II}(\eps_1,\eps_2; \omega)$ for which (Bethe-Salpeter-like) equations have been derived. In constructing these equations the analytic structure of  $\Gamma^{I/II}(\eps_1,\eps_2; \omega)$ is essential and it was discussed in detail. Here we take simpler approach which allows for a more transparent discussion. We note that all major results can also be obtained by continuing the line of analysis in Ref.~\onlinecite{MFL}. 

We start by calculating each term in the series for  $Z^{\alpha}_{\omega}$ in Eq.~(\ref{eq:Zalpha}). We introduce a function  $Z^{\alpha}_{\omega} (\eps)$ which is obtained by omitting the last section and the corresponding energy sum in each term in Eq.~(\ref{eq:Zalpha}); we define  $Z_{\omega}(\eps) = \sum_{n=0..\infty}Z_{\omega}^{(n)} (\eps)$ (to avoid excessive indexing we omit $\alpha$ index where it does not lead to confusion). Zero term is one
\begin{align}\label{eq:aaa}
Z^{(0)}_{\omega}(\eps) = 1
\end{align}
The first term is 
\begin{align}\label{eq:aaa}
&Z^{(1)}_{\omega} (\eps_2)
=g_{\alpha}\sum_{i\eps_1} \lefttriang\verbare{\eps_1+\omega}{\eps_1}{}{\eps_2+\omega}{\eps_2}  
\end{align}
The function $Z_{\omega}(\eps)$ has only three analytic pieces in the complex plane of $\eps_2$; these are separated by branch cuts  at $Im\eps_2=0$ and $Im\eps_2=-\omega_n$, i.e., its analytic structure is equivalent to that of the particle hole section which describes free propagation of particle-hole excitation. To label the analytic pieces of $Z_{\omega}(\eps)$ we use notation which is commonly used for particle-hole section~: 
\begin{align}\label{eq:aaa}
AA \rightarrow&\qquad  \Im\eps<-\omega_n \notag\\
RA \rightarrow&\qquad  -\omega_n < \Im\eps<0 \notag\\
RR \rightarrow&\qquad  0<\Im\eps \,.
\end{align}
We have assumed that $\Im\omega>0$, i.e., we are calculating retarded part of the response function. The main advantage of working with $Z_{\omega}(\eps)$ is that in the RA range of $i\eps_n$ it has only one analytic piece.  Compare this with the full rescattering amplitude $\Gamma^{II/I}$ analyzed in Ref.~\onlinecite{MFL} which has two analytic pieces in the RA range of frequencies. 

The analytic continuation for the RA analytic piece has a structure 
\begin{multline}\label{eq:Zepsilon-1}
Z^{(1)}_{\omega} (\eps_2)^{RA}\\
=-g_{\alpha}\int d\eps_1 \Bigg\{-\frac12\tanh\frac{\eps_1}2 \verbare{}{}{II}{}{}  
+\frac12\tanh\frac{\eps_1+\omega}2 \verbare{}{}{I}{}{}\\
+\frac12\coth\frac{\eps_1-\eps_2}2 \verbare{}{}{II\!-\!I}{}{} 
\Bigg\} 
\end{multline}
Here and in rest of discussion we omit temperature in the trigonometric factors where it does not lead to confusion. Note that the amplitude for the rescattering event (given by the vertical wiggly line), viewed as a function of $\eps_1,\eps_2$, has the same analytic structure as the fully renormalized amplitude   $\Gamma^{I/II}(\eps_1,\eps_2; \omega)$; since only one rescattering event is involved here, the analytic pieces $II/I$ directly correspond to  $R/A$ analytic pieces of $\mode(\omega)$. The first and second terms in Eq.~(\ref{eq:Zepsilon-1}) are generated by the branch cuts at $Im\eps_1=0$ and $Im\eps_1=-\omega_n$. The third term originates in the dynamic mode which introduces a branch cut at $Im\eps_1=\Im\eps_2$. We rearrange Eq.~(\ref{eq:Zepsilon-1}) in terms of $\mode^+\equiv\mode^{II}+\mode^{I}=2\Re\mode$ and $\mode^-\equiv\mode^{II}-\mode^{I}=2i\Im\mode$ 
\begin{multline}\label{eq:Z1}
Z_{\omega}^{(1)}(\eps_2) _{RA}=  g_{\alpha} \int d\eps_1\Bigg\{ -\frac14 \scriptstyle{\Big[-\tanh\frac{\eps_1}2+\frac12\tanh\frac{\eps_1+\omega}2 \Big]}
\verbare{}{}{II+I}{}{}\\ 
-\frac14 \scriptstyle{\Big[-\tanh\frac{\eps_1}2-\tanh\frac{\eps_1+\omega}2+
2\coth\frac{\eps_1-\eps_2}2 \Big]}
\verbare{}{}{II\!-\!I}{}{}  \Bigg\}
\end{multline}
In this form the energy integration is restricted to the vicinity of the Fermi surface in both terms. The momentum integration in the RA particle-hole section is done by deforming contour in the complex plane of momentum represented by an energy variable $\xi$ defined via $\nu\xi=\eps_{|\bold{p}+\delta\bold{p}|}$. We define a function $S(\omega)_{RA}$ 
\begin{multline}\label{eq:phsection}
2\pi\nu i S(\eps;\omega,q)_{RA} =
\int\frac{d^2p}{(2\pi)^2} G_{\mathbf{p+q}}^R(\eps+\omega) G_{\mathbf{p}}^A(\eps)\\
= 2\pi i\nu \int\frac{d\theta}{2\pi} \frac1{\omega-qv_F\cos\theta-[\Sigma^R(\eps+\omega)-\Sigma^A(\eps)]}
\end{multline}
In the complex plane of $\xi$  the poles of two Green's functions in the RA section are on different sides of the contour of integration.  the distance of the pole from the real axis is proportional to the imaginary part of selfenergy with the coefficient $1+d\Re\Sigma/dk$. Since in cuprates the momentum dependence of selfenergy is very weak, when imaginary part of selfenergy is small, the pole in the complex plane of $\xi$ is close to the real axis. This is the Migdal theorem\cite{Pitaevskii} applied to cuprates.  $\xi$-integration extends to a large momenta,  $\xi\sim\eFermi$, therefore for $\eps,\omega$ close to Fermi surface the momentum integration can be done by taking a residue at the pole on one of the sides of the real axis (the contour pinching). Though it is not obvious that in cuprates (where the Green's function has a singular selfenergy) analytic structure of the particle-hole section in the complex plane of $\xi$ is similar to the conventional case, the calculation path taken here is consistent because we make sure that the conservation laws are obeyed (see below). For non-spherical Fermi surface Eq.~(\ref{eq:phsection}) is replaced with 
\begin{multline}\label{eq:S-angles}
2\pi\nu_0 i S(\eps;\omega,q)^{\alpha} =
\int\frac{d^2p}{(2\pi)^2} F_{\alpha}^2(\bm{p}) G_{\bm{p+q}}^R(\eps+\omega) G_{\bm{p}}^A(\eps)\\
= 2\pi i \nu_0 \intA{\theta}\frac{F_{\alpha}^2(\theta)}{\omega-qv_F\cos\theta-[\Sigma(\eps+\omega)_{\theta}^R-\Sigma(\eps)_{\theta}^A]}
\end{multline}

Analytic continuation in RR and AA analytic pieces of $Z_{\omega}(\eps)$ leads to an expression in which the energy integration is no more restricted to the vicinity of the Fermi surface. The combined energy and momentum integration is logarithmically divergent,  compare this discussion with the one following Eq.~(\ref{eq:higherterms}). However there is no significant contribution to the non-analytic behavior near the Fermi surface since both poles in the complex plane of $\xi$ are on the same side of the contour of integration; we only consider RA analytic piece in the ladder (and omit RA subscript). 

To keep discussion simple and to illustrate the main point we first restrict ourselves to small frequencies, $\omega\ll T$ and omit all terms that vanish in this limit; this means we omit the real part of the selfenergy  and the first term in Eq.~(\ref{eq:Z1}). It will be trivial to restore complete equations later on, see discussion leading to Eqs.~(\ref{eq:EquationY})~and~(\ref{eq:EquationZ}). We write 
\begin{multline}\label{eq:aaa}
Z_{\omega}^{(1)}(\eps_2) = g_{\alpha}\int d\eps_1 \scriptstyle{\Big\{ -\frac12 \Big[-\tanh\frac{\eps_1}2+
\coth\frac{\eps_1-\eps_2}2 \Big]\Big\}}
\verbare{}{}{II\!-\!I}{}{}  \\
= -g_{\alpha}\int d\eps_1 \frac12[\coth\frac{\eps_1-\eps_2}2 - \tanh\frac{\eps_1}2 ]
S(\eps_1)\mode^-(\eps_1-\eps_2)\\
= \int d\eps_1  \frac{2ig_{\alpha}}{\omega+ 2ig_0\eps\coth\frac{\eps_1}2} 
\frac12[1 - \tanh\frac{\eps_1-\eps_2}2\tanh\frac{\eps_1}2 ]
\end{multline}
where 
\begin{align}\label{eq:aaa}
\mode^-(\omega) =& -2i\tanh\frac{\omega}{2T}
\end{align}
and
\begin{align}\label{eq:aaa}
S(\eps) =& \frac1{\omega+2ig_0\eps\coth\frac{\eps}2 }
\end{align}
Now consider the next term in the series
\begin{align}\label{eq:aaa}
Z_{\omega}^{(2)}(\eps) = g_{\alpha}^2\lefttriang\verbare{}{}{}{}{}\!\!\!\!\!\!\verbarestep{}{}{}
\end{align}
To make analytic continuation for the frequency sum in the second section we have to consider the analytic structure to the left, $Z^{(1)}(\eps)$,  and to the right, $\mode^-(\eps_2-\eps_3)$. It has been noted that  $Z_{\omega}(\eps)$  has only one analytic piece in the RA range of Matsubara frequencies. We only have to consider a branch cut at $\Im\eps_2=\Im\eps_3$ which originates in the second rescattering event and therefore the analytic continuation for the frequency sum is done the same way as in the case of  $Z^{(1)}(\eps)$. Keeping only terms that do not vanish in the limit  $\omega \rightarrow0$, we write
\begin{multline}\label{eq:aaa} 
Z^{(2)}(\eps_3) \\
= \int d\eps_1d\eps_2  
\frac{2ig_{\alpha}}{\omega+ 2ig_0\eps_1\coth\frac{\eps_1}2} 
\frac12[1 - \tanh\frac{\eps_1-\eps_2}2\tanh\frac{\eps_1}2 ]\\
\times\frac{2ig_{\alpha}}{\omega+ 2ig_0\eps_2\coth\frac{\eps_2}2} 
\frac12[1 - \tanh\frac{\eps_2-\eps_3}2\tanh\frac{\eps_2}2 ]
\end{multline}
The rest of the terms $Z^{(n)}_{\omega}(\eps)$ are constructed the same way. The form of each term in the series  $Z^{(n)}_{\omega}(\eps)$ indicate that the sum  $Z_{\omega}(\eps)$ satisfies an equation
\begin{align}\label{eq:Z-epsilon-omega}
&Z_{\omega}(\eps) = 1 + \notag\\
&\int d\eps' Z_{\omega}(\eps') \frac{2ig_{\alpha}}{\omega+ 2ig_0\eps'\coth\frac{\eps'}2} 
\frac12[1 - \tanh\frac{\eps'-\eps}2\tanh\frac{\eps'}2 ]  
\end{align}
To obtain $Z_{\omega}$ we have to add the last section and perform the frequency sum. Here the analytic continuation in the frequency sum leads to a standard trigonometric factor: in the RA section there are no additional branch cuts. We obtain  
\begin{multline}\label{eq:Z-omega}
Z_{\omega} = -\int d\eps Z_{\omega}(\eps) 
\frac{1}{\omega+ 2ig_0\eps\coth\frac{\eps}2} 
\frac12[\tanh\frac{\eps+\omega}2-\tanh\frac{\eps}2 ] \\
=-\int d\eps Z_{\omega}(\eps) 
\frac{1}{\omega+ 2ig_0\eps\coth\frac{\eps}2}  
\left[\omega \frac{d \tanh\frac{\eps}2}{2d\eps}\right]  
\end{multline}
the minus sign is a result of our definition of $Z_{\omega}(\eps)$ which starts at $Z_{\omega}^{(0)}(\eps)=1$; to check the sign consider $g_{\alpha}=0$ in which case calculation of $Z _{\omega}$ reduces to a single bubble diagram with the result $-1$. In the second  line  of Eq.~(\ref{eq:Z-omega}) we have used 
\begin{align}\label{eq:aaa}
\frac12\Bigg[\tanh\frac{\eps+\omega}2 -\tanh\frac{\eps}2\Bigg] \xrightarrow[\omega \rightarrow0]{}
\omega \frac{d\tanh\frac{\eps}2}{2d\eps}  
\end{align}

\subsection{\label{sec:conservation} particle number conservation.}

Particle conservation requires that the density correlation function vanishes identically for $q=0$; it is therefore necessary that in this limit the dynamic part of the density correlation function approach real, $\omega$-independent value. With our definition of the singular segment we have to check that    $Z_{\omega}^{\alpha=A_{1g}} =-1$ if calculation is made in the density channel, $A_{1g}$;  see Refs.~\onlinecite{Leggett,MFL} for details. We start with Eqs.~(\ref{eq:Z-epsilon-omega})~and~(\ref{eq:Z-omega}) which have been derived for $\omega\ll T$. Eq.~(\ref{eq:Z-omega}) has an overall factor $\omega$; in the limit $\omega\rightarrow0$ the function  $Z_{\omega}$ will vanish unless $Z_{\omega}(\eps)$ is singular; we conclude that to ensure particle conservation function $Z_{\omega}(\eps)$ has to behave as  $Z_{\omega}(\eps)\propto1/\omega$ at small omega. To see that this is indeed the case, we observe that at finite temperature the denominator of the particle-hole section $S(\omega)$ is finite and the frequency sum in Eq.~(\ref{eq:Z1}) has a contribution which does not vanish in the limit $\omega \rightarrow0$. Therefore each diagram  $Z^{(n)}_{\omega=0}(\eps)$ is finite (and positive) and at $\omega=0$ the sum  $Z_{\omega=0}(\eps)=\sum Z^{(n)}_{\omega=0}(\eps) $ will diverge. We expect that the effect of a small but finite $\omega$ is to regularize the sum with the result $Z_{\omega}(\eps)\propto1/\omega$. 

We see that the fact that  $Z^{(n)}_{\omega=0}(\eps)$ is finite in the limit of $\omega=0$ is crucial for the consistency of our calculation. The trigonometric factor in the first term of Eq.~(\ref{eq:Z1}) vanishes in the limit of $\omega\rightarrow0$ while the second remains finite. Physically, this nonvanishing contribution originates from the singularity in the dynamic mode, the same singularity which is responsible for the inelastic behavior of the quasiparticle selfenergy. There is a parallel with the calculation of diffusion ladder for the elastic disorder; in the disorder ladder the absence of a factor of $\omega$ in each particle-hole section is natural because there is no frequency sum in the intermediate sections. In the case of dynamic mode the finite value (in the limit $\omega \rightarrow0$) of imaginary part of selfenergy (``outs'') is compensated by the finite result of the frequency sum in the particle-hole section (``ins''). Similar to elastic disorder, their complete cancelation in the calculation of $Z_{\omega}^{\alpha=A_{1g}}$ ensures particle conservation. Unlike disorder, it does not seem possible to calculate each term  $Z^{(n)}_{\omega}(\eps)$ separately and follow the cancellation explicitly.

We now observe that Eq.~(\ref{eq:Z-epsilon-omega}) has a closed solution in the density channel~: 
\begin{align}\label{eq:Z-density}
Z_{\omega}(\eps) = \frac{2ig_0}\omega \times \frac{\omega+ 2ig_0\eps\coth\frac{\eps}2}{2ig_0}  
\end{align}
To check, we substite this expression into the right hand side of Eq.~(\ref{eq:Z-epsilon-omega}) 
\begin{align}\label{eq:aaa}
&Z_{\omega}(\eps) = 1 + \frac{2ig_{0}}{\omega}\int d\eps' 
\frac12[1 - \tanh\frac{\eps'-\eps}2\tanh\frac{\eps'}2 ]  
\end{align}
and use the fact that the integral here has the same form as the one which appears in the calculation of selfenergy, see  Eq.~(\ref{eq:Im-Sigma})
\begin{align}\label{eq:recall}
\Sigma^-(\eps) =& g_0\int d\eps' \mode^-(\eps'-\eps)
\frac12[\coth\frac{\eps'-\eps}2-\tanh\frac{\eps'}2 ]  
\notag\\
=& -2ig_0 \eps\coth\frac{\eps}2\,.
\end{align}
We have 
\begin{align}\label{eq:aaa}
Z_{\omega}(\eps) = 1 + \frac{2ig_0 \eps\coth\frac{\eps}2}{\omega} 
\end{align}
which coincides with Eq.~(\ref{eq:Z-density}). Substituting Eq.~(\ref{eq:Z-density}) into Eq.~(\ref{eq:Z-omega}) we obtain 
\begin{align}\label{eq:aaa}
Z_{\omega}^{\alpha=A_{1g}} = -1
\end{align} 
which proves particle conservation. 

In the rest of this section we extend the above analysis for arbitrary $\omega/T$ and restore the real part of the selfenergy (and the first term in Eq.~(\ref{eq:Z1})). First, we rewrite Eq.~(\ref{eq:Z-epsilon-omega}) in the form which make the cancelation more transparent. Introduce a new variable $Y^{\alpha}_{\omega}(\eps)$ which corresponds to  adding a particle-hole section, $S(\epsilon)$ to each diagram in $Z^{(n)}_{\omega}(\eps)$ 
\begin{align}\label{eq:distribution}
 Y_{\omega}^{\alpha}(\eps) =& Z_{\omega}^{\alpha} (\eps)
\frac{1}{\omega+ 2ig_0\eps\coth\frac{\eps}2} \notag\\
Z_{\omega}^{\alpha} =&-\int d\eps {Y_{\omega}^{\alpha} (\eps)} 
\left[\omega \frac{d \tanh\frac{\eps}2}{2d\eps}\right] 
\end{align}
$Y_{\omega}(\eps)$ satisfies an equation 
\begin{multline}\label{eq:aaa}
[{\omega} + 2ig_0\eps\coth\frac{\eps}2] Y_{\omega}^{\alpha}(\eps) 
= 1 \\
+2ig_{\alpha}\int d\eps' Y_{\omega}^{\alpha}(\eps') 
\frac12[1 - \tanh\frac{\eps'-\eps}2\tanh\frac{\eps'}2 ]  
\end{multline}
We use Eq.~(\ref{eq:recall}) to rewrite it as 
\begin{multline}\label{eq:aaa23}
{\omega} Y_{\omega}^{\alpha} (\eps) = 1 \\
+\int d\eps' \Big[2ig_{\alpha}Y_{\omega}^{\alpha} (\eps')-2ig_{0}Y_{\omega}^{\alpha} (\eps)\Big] 
\frac12[1 - \tanh\frac{\eps'-\eps}2\tanh\frac{\eps'}2 ]  
\end{multline}
In this form it is obvious that in the density channel, $g_{\alpha}=g_0$, the solution of this equation is independent of $\eps$ and therefore has a form
\begin{align}\label{eq:solutionYdensity}
Y_{\omega}(\eps) = \frac{1}{\omega}
\end{align}
When substituted in Eq.~(\ref{eq:distribution}) it gives $Z_{\omega}=-1$. 

To obtain an equation for $Y(\eps)$ which is valid for arbitrary  $\omega/T$ we start with Eq.~(\ref{eq:Z1}) and repeate the same sequence of mathematical steps. The result is
\begin{multline}\label{eq:EquationY}
\omega Y_{\omega}(\eps)^{\alpha} = 1 \\
+ \int d\eps' \Big[g_{\alpha}Y_{\omega}^{\alpha}  (\eps')-g_{0}Y_{\omega}^{\alpha} (\eps)\Big] \mode_{\omega}^K(\eps'-\eps)
\end{multline}
which together with 
\begin{align}\label{eq:EquationZ}
& Z_{\omega}^{\alpha} = -\int d\eps {Y_{\omega}^{\alpha} (\eps)} 
\frac12[\tanh\frac{\eps+\omega}2-\tanh\frac{\eps}2 ] 
\end{align}
determines the singular segment for arbitrary  $\omega$ and $T$. Here we have defined a function 
\begin{multline}\label{eq:Tkernel}
\mode_{\omega}^K(\eps'-\eps) = \\
\mode^-(\eps'-\eps)\frac12 \Big[\coth\frac{\eps'-\eps}2 - \frac12\big[\tanh\frac{\eps'+\omega}2 + \tanh\frac{\eps'}2 \big]\Big] \\
+\mode^+(\eps'-\eps)\frac14\Big[\tanh\frac{\eps'+\omega}2 - \tanh\frac{\eps'}2 \Big]
\end{multline}
In the density channel the solution of Eq.~(\ref{eq:EquationY}) is independent of the temperature and coincides with Eq.~(\ref{eq:solutionYdensity}). This concludes the proof of a particle conservation property within the framework of our mathematical discussion.  Eqs.~(\ref{eq:EquationY})~and~(\ref{eq:EquationZ}) are the main technical result of our discussion; for the purpose of the analysis of the response functions on the Fermi surface they replace ladder equations derived in Ref.~\onlinecite{MFL}. 

\subsection{Calculation of $Z^{\alpha}_{\omega}$ in Raman channels, $g_{\alpha}\neq g_0$.}

In Raman channels, $g_{\alpha}\neq g_0$, the solution of Eq.~(\ref{eq:EquationY}) is a function of $\eps$ and it appears that closed solution of the equation is not possible. Here we obtain approximate solution which is consistent with the the analytic structure of the singular segment in Raman channels. We solve Eq.~(\ref{eq:EquationY}) for  $\omega\ll T$ and $\omega\gg T$ and give and interpolating formula which bridges the two cases. Analysis of experimental data will require more rigorous analysis of Eq.~(\ref{eq:EquationY}). First consider the case of large frequencies, $\omega\gg T$. We omit the $\mode^+$ part in the expression for $\mode^K$; Eqs.~(\ref{eq:EquationY})~and~(\ref{eq:EquationZ}) take a form 
\begin{align}\label{eq:Equation-large-omega}
&\omega {Y_{\omega}^{\alpha} (\eps)} = 1 + {2i}\int\limits_{\min\{-\omega,0\}}^{\max\{-\omega,0\}} d\eps' \Big(g_{\alpha}Y_{\omega}^{\alpha} (\eps')-g_{0}Y_{\omega}^{\alpha} (\eps)\Big) 
\notag\\
& Z_{\omega}^{\alpha} = -\sign{\omega}\int\limits_{\min\{-\omega,0\}}^{\max\{-\omega,0\}}  
d\eps {Y_{\omega}^{\alpha} (\eps)} 
\end{align}
In the first equation we have used the fact that  $-\omega<\eps<0$, see Eq.~(\ref{eq:EquationZ}). The solution $Y(\eps)$ is independent of $\eps$~; the function $Y^{\alpha}_{\omega}$ satisfies an equation  
\begin{align}\label{eq:EquationYraman}
\omega Y_{\omega}^{\alpha}= 1 + 2 i (g_{\alpha}-g_0) |\omega| Y_{\omega}^{\alpha} 
\end{align}
We find 
\begin{align}\label{eq:ZsmallT}
Z_{\omega}^{\alpha} = \frac{-\omega}{\omega + 2 i (g_0 - g_{\alpha})|\omega|}
\end{align}
With this expression in mind we can write singular segment in the form
\begin{align}\label{eq:aaa}
 Z_{\omega}^{\alpha} = \frac{-\omega}{\omega + 2 \big[g_0 - g_{\alpha}\big] \Psi(\omega)_T }
\end{align}
where $\Psi(\omega)_T$ is analytic in the upper half of complex $\omega$-plane (i.e., it is a retarded function, same as $Z(\omega)$). Comparing with   Eq.~(\ref{eq:ZsmallT}) (and using results in the rest of this section) we write 
\begin{align}\label{eq:ImPsiTzero}
\Im\Psi(\omega)_T = 
\begin{cases}
|\omega|,  &  \qquad \omega\gg T\\
c_0 T  , &  \qquad \omega\ll T
\end{cases} 
\end{align}
where $c_0\approx2.3$. Note that analyticity of $Z^{\alpha}(\omega)$ in the upper half of the complex plane is guaranteed by the fact that  $g_0 - g_{\alpha}>0$.  Having in mind our experience with the calculation of the selfenergy we write an interpolating expression 
\begin{align}\label{eq:aaa}
\Im\Psi(\omega)_T =   \smear{p} \Im\Psi(\omega-s)_T
\end{align}
Where constant $p$ is fixed by the behavior of $\Psi(\omega)$ at $\omega\ll T$; we find $p\approx2\log2/2.3\approx0.59$. The real part of $\Psi(\omega)$ is calculated from the first line in Eq.~(\ref{eq:ImPsiTzero}) via Kramers-Kronig integral. We find 
\begin{align}\label{eq:aaa}
&  \Im\Psi(\omega)_{T=0}  = |\omega|, \quad  \Re\Psi(\omega)_{T=0} = \omega \log\frac{\Lambda}{|\omega|} \notag\\
& \Psi(\omega)_T = \smear{p} \Psi(\omega-s)_{T=0}
\end{align}
The cutoff $\Cutoff_{\Psi}$ has to be of order of the upper cutoff $\Cutoff_0$ in the dynamic mode, see Eq.~(\ref{eq:MFL-chi}). 

In the rest of this Section we analyze Eqs.~(\ref{eq:EquationY})~and~(\ref{eq:EquationZ})  in the limit $\omega\ll T$ which determines constant $c_0$ in Eq.~(\ref{eq:ImPsiTzero}). Eqs.~(\ref{eq:EquationY})~and~(\ref{eq:EquationZ}) take a form 
\begin{align}\label{eq:Equation-small-omega}
&\omega{Y_{\omega}^{\alpha} (\eps)} = 1 \notag\\
&+ 2i\int d\eps' \Big(g_{\alpha}Y_{\omega}^{\alpha} (\eps')-g_{0}Y_{\omega}^{\alpha} (\eps)\Big)  \frac12\Big[1 - \tanh\frac{\eps'-\eps}2 \tanh\frac{\eps'}2 \Big] 
\notag\\
& Z_{\omega}^{\alpha} = -\omega\int d\eps {Y_{\omega}^{\alpha} (\eps)} 
\frac{d \tanh\frac{\eps}2}{2d\eps}
\end{align}
Introducing new variables 
\begin{align}\label{eq:aaa}
\eta=\tanh\frac{\eps}2 ,\qquad \eta'=\tanh\frac{\eps'}2\,.
\end{align}
we rewrite Eq.~(\ref{eq:Equation-small-omega}) in the form
\begin{align}\label{eq:aaa21}
&\omega{Y_{\omega}^{\alpha} (\eta)} = 1 
+ {2i}\int_{-1}^1d\eta'\frac{\big[g_{\alpha}Y_{\omega}^{\alpha} (\eta')-g_{0}Y_{\omega}^{\alpha}(\eta)\big]}{1 - \eta'\eta} \notag\\
& Z_{\omega}^{\alpha} = -\omega\int_{-1}^1\frac{d\eta}2 Y_{\omega}^{\alpha}(\eta)
\end{align}
Multiplying the first equation by $\int d\eta/2$ we write
\begin{align}\label{eq:aaa}
\int_{-1}^1\frac{d\eta}2\Big[ \omega + {2i}\big(g_0 - g_{\alpha}) \frac{2\atanh\eta}{\eta}\Big] Y_{\omega}^{\alpha} (\eta) = 1 
\end{align}
where we have used $\int_{-1}^1d\eta'/(1-\eta\eta') = (2\atanh\eta)/\eta$. The solution is 
\begin{align}\label{eq:aaa22}
Y_{\omega}^{\alpha} (\eta) = \frac{1+c(\eta)}{ \omega + {2i}\big(g_0 - g_{\alpha}) \frac{2\atanh\eta}{\eta}}
\end{align}
where $c(\eta)$ is a function which has zero mean, $\int_{-1}^1c(\eta)=0$; it can be determined by substituting Eq.~(\ref{eq:aaa22}) into Eq.~(\ref{eq:aaa21}); for our approximation we take $c(\eta)=0$. Substituting this into Eq.~(\ref{eq:aaa21}) and taking the limit $\omega \rightarrow0$ we find $Z^{\alpha}(\omega) = -\omega/[-2i(g_{\alpha}-g_0) c_0 T]$ where $1/c_0 = \int_{-1}^1 \eta d\eta/(4\atanh\eta) \approx0.43$ and therefore $c_0 \approx 2.3$.  

\begin{acknowledgments}
Numerous discussions with Chandra Varma, Vivek Aji, Alexander Finkel'stein and Lev Gor'kov are greatly acknowledged. Useful comments by Lev Bulaevskii are appreciated. It is a pleasure to acknowledge the hospitality of Kavli Institute of Theoretical Physics and Los Alamos National Lab where parts of this work were done.
\end{acknowledgments}

\end{document}